\documentclass[preprint]{aastex}
\slugcomment{{\sc Submitted to ApJ:}}
\usepackage{float}

\newcommand{\msun}{\ensuremath{\rm{M}_\odot}}
\newcommand{\rsun}{\ensuremath{\rm{R}_\odot}}
\newcommand{\Lsun}{\ensuremath{\rm{L}_\odot}}

\newcommand{\sneia}{SNe Ia}
\newcommand{\snia}{SN Ia}

\begin{document}

\title{Type Ia Single Degenerate Survivors Must Be Overluminous}  
\shorttitle{Type Ia Single Degenerate Survivor} 
\shortauthors{Shappee, Kochanek \& Stanek}

\author{
{Benjamin J. ~Shappee}\altaffilmark{1},
{C. S. ~Kochanek},
and
{K. Z. ~Stanek}
}

\affil{Department of Astronomy, 
The Ohio State University, Columbus, Ohio 43210, USA}

\email{shappee@astronomy.ohio-state.edu, ckochanek@astronomy.ohio-state.edu, kstanek@astronomy.ohio-state.edu
}

\altaffiltext{1}{NSF Graduate Fellow}

\date{\today}

\begin{abstract}

In the single-degenerate (SD) channel of a Type Ia supernovae (SN Ia)
explosion, a main-sequence (MS) donor star survives the explosion but it is stripped of mass and shock heated.  An essentially unavoidable consequence of mass loss during the explosion is that the companion must have an overextended envelope after the explosion.  While this has been noted previously, it has not been strongly emphasized as an inevitable consequence. We calculate the future evolution of the companion by injecting $2\--6 \times 10^{47}\;$ergs into the stellar evolution model of a $1$ \msun{} donor star based on the post-explosion progenitors seen in simulations.  We find that, due to the Kelvin-Helmholtz collapse of the envelope, the companion must become significantly more luminous
($10 \--10^3\;\Lsun$) for a long period of time ($10^3\--10^4\;$years).
The lack of such a luminous ``leftover'' star in the LMC supernova
remnant SNR $0609-67.5$ provides another
piece of evidence against the SD scenario. We also show that none of the stars
proposed as the survivors of the Tycho supernova, including Tycho G,
could plausibly be the donor star. Additionally, luminous donors closer than $\sim 10\;$Mpc 
should be observable with the Hubble Space Telescope starting $\sim 2$ years post-peak.  Such systems include SN 1937C, SN 1972E, SN 1986G, and SN
2011fe. Thus, the SD channel is already ruled out for at least two nearby \sneia{} and can be easily tested for a number of additional ones. We also discuss similar implications for the companions of core-collapse SNe.

\end{abstract}

\section{Introduction}
\label{sec:introduc}

Type Ia supernovae (\sneia{}) are among the most luminous explosive
events known, and are observable from halfway across the Universe
(e.g., Rodney et al. 2012).  The peak luminosities of \sneia{} are well
correlated with the decline rate of their light curves, thereby permitting their use as standardizable candles \citep{phillips93, hamuy95,
riess95}.  The combination of these two properties has made \sneia{} very useful for measuring cosmological parameters \citep{riess98, perlmutter99}.  As surveys have grown in scope, systematic errors are becoming the
dominant source of uncertainty in Type Ia SNe-based cosmological constraints
\citep{wood-vasey07, kessler09, guy10, conley11}.

Even as \sneia{} have been crucial in shaping our understanding of
the universe, the nature of their progenitor systems remains
theoretically ambiguous and observationally elusive (for a review see
\citealp{wang12}). While it is commonly accepted that \sneia{} result
from the thermonuclear explosion of a carbon-oxygen white dwarf (WD)
in a close binary system, the nature of the binary companion and the
chain(s) of events leading up to the SN explosion are still uncertain.
Broadly speaking, there are two dominant progenitor models.  The first
is the double degenerate (DD) scenario where the companion is also a
WD \citep{tutukov79, iben84, webbink84}.  The second is the single degenerate
(SD) scenario where the companion is a non-degenerate object: a main
sequence (MS) star, a red giant (RG), a sub-giant, or a He star
\citep{whelan73, nomoto82}.  While the SD scenario is widely considered to
be the more plausible channel for \sneia{}, there is mounting
evidence suggesting that this view is incorrect. First, \sneia{} progenitor models predict outflows during the presupernova evolution creating windblown cavities, however, \citet{badenes07} find that the X-ray observations of 7 young \snia{} remnants are inconsistent with this picture\footnote{One notable exception, RCW 86, is discussed by \citet{williams11}.}.  Second, it has recently been shown that rate of WD mergers in the Galactic disk is comparable to that of the \sneia{} rate in Milky Way like galaxies, seemingly in support of the DD scenario \citep{badenes12}. Additionally, giant
donors seem to be ruled out observationally as the dominant channel\footnote{There are, however, a number of systems which are considered to be candidates for the  symbiotic channel (e.g. U Sco, RS Oph and TCrB; \citealp{parthasarathy07}).} (e.g., \citealp{mattila05, leonard07, hayden10, bianco11,
brown12b, li11, bloom12, brown12, edwards12, schaefer12}), and in this study we do not consider them as possible progenitor systems. An He star donor was recently ruled out for the nearby Type Ia SN~2011fe \citep{li11}, while main sequence and sub-giant companions have mounting evidence against them (e.g., \citealp{leonard07, brown12b, bloom12, schaefer12}), but might still remain a significant channel for the production of \sneia{}.   

One approach to constraining the SD channel is to examine the effects of the explosion on the donor star \citep{colgate70, cheng74, wheeler75,
fryxell81, taam84, livne92, marietta00, meng07, pakmor08,
pan12}. While many aspects of the interaction of a Type Ia
explosion with the donor star, such as its kick velocity
(e.g., \citealp{ruiz-lapuente04}), early-time shock emission
(e.g., \citealp{kasen10, hayden10, bianco11, brown12}), the statistical properties of the surviving companion population \citep{han08, wang10}, and the amount of material removed by the
explosion \citep{mattila05, leonard07}, have received close attention,
little has been made of the dramatic changes in the properties of the donor
following the explosion.
\citet{pan12} recently emphasized that fully $65 \%$ of the mass removed from the companion is due to ablation (shock-mediated heat transfer), not stripping (momentum transfer),\footnote{
This result contradicts their previous work, \citet{pan10}, which used the mixing of the SN ejecta at late-times in 2D simulations and not tracer particles in 3D simulations to estimate the ratio of ablated-to-stripped material. As discussed in \citet{pan12}, this 2D method misclassified material that was ablated early in the simulation as having been stripped.} 
so mass cannot be removed without heating the star because ablation and heating are not independent variables. Thus, as mass is ablated from the companion, the entire stellar interior is also shock heated and the remaining envelope is significantly puffed up \citep{marietta00}. 
\citet{podsiadlowski03} modeled the future evolution of
\snia{} companions by considering a constant amount of mass stripped
from the donor star while treating the energy injected into the
envelope of the companion as a free parameter, and varied it almost by an
order of magnitude.  While not favored by \citet{podsiadlowski03}, models with small amounts of energy injected into the remaining star compared to the binding energy of the stripped material could actually become less luminous than the star's luminosity prior to the explosion.  These low luminosity models are not, however, consistent with the simulations.  

In this paper we focus on the post-explosion state of the SD donor star.  They are required to be more luminous than before the explosion by the physics observed in simulations of shock interactions.  As a result, the
recent observations of nearby SN remnants (SNRs) to detect or constrain the surviving star are far more stringent than generally assumed. In
Section \ref{sec:PostProp}, we model the evolution of the donor star following the interaction with the SN ejecta and compare our results to previous discussions. In Section \ref{sec:Individual}, we investigate nearby
systems individually, first looking at SNRs in the Galaxy and the LMC and then discussing
nearby extragalactic \sneia{}.  Section \ref{sec:core} briefly discusses the effect of shock interactions on the binary companions of core-collapse SNe.  In Section
\ref{sec:conclusion} we review our conclusions.

\section{Post-Explosion Properties of the Secondary}
\label{sec:PostProp}

There have been several numerical studies of the impact of the SN ejecta on a companion star of a \snia{}.  \citet{marietta00}
performed two-dimensional simulations of the impact of the SN ejecta
on MS stars, sub-giants, and RGs using an Eulerian hydrodynamics code
and found that MS and sub-giant donors lost $\sim 15\%$ of their
mass as a result of the impact.  \citet{meng07} then pointed out that
the pre-supernova mass transfer required by the SD model would change
the stellar structure of the companion, causing it to become more
compact than an isolated MS star.  From this they concluded that less mass should be removed from the donor envelope than
predicted by the models of \citet{marietta00}.  \citet{pakmor08} 
confirmed this using a three-dimensional smoothed particle
hydrodynamics (SPH) code to model the impact of SN ejecta on a
companion whose structure was specified by the binary evolution study
of \citet{ivanova04}.  Finally, \citet{pan12} performed the most
detailed study to date, with the inclusion of symmetry-breaking effects
(i.e. orbital motion, rotation of the companion, and Roche-lobe
overflow) in a simulation using the multi-dimensional adaptive mesh
refinement code FLASH.  They found that these effects again raise the
mass unbound from the secondary to $16\%$ for a MS donor.

These studies show that there is a generic sequence of events during the impact of
the SN ejecta on the companion.  The initial impact sends a
shock into the donor's stellar envelope which also sends a reverse
shock back into the SN shell and creates a contact discontinuity between
the two. The reverse shock develops into a bow shock around the
companion which deflects much of the SN ejecta around the donor.
Meanwhile, the forward shock penetrates the companion and the center of the
shock slows as it encounters the steep density gradient of the stellar
interior, accelerating once more after passing through the core.  The
wings of the shock, which propagate through the outer envelope, move faster
than they do through the center because they are propagating through regions of lower
density, causing the shock to become highly curved.  The curved shock
meets itself at the backside of the companion, depositing much of its
energy in the donor's envelope and ablating significant amounts of mass from the surface.  This creates a back pressure on the rest of the companion,
decreasing the kick imparted by the SN blast wave.  After most of the
SN ejecta has passed the donor, the shock-heated outer envelope of
the companion begins to expand and some of it becomes unbound.  The majority
of the unbound material is at low velocities and is mixed with the inner
iron-group layers of the SN ejecta. Even though some material is
unbound from the companion, the majority of its mass remains bound to the star.  The net heating of the star is not reported, but in the standard $1$ \msun{} MS model of \citet{marietta00} the star is left with a hot, extended, asymmetric envelope containing $\sim 10\%$ of the mass, while the central pressure is halved.  There is almost no discussion of the subsequent appearance of the stars other than in \citet{marietta00}.  They note that its core nuclear reaction rates will be diminished for a time-scale of $10^{3} \-- 10^{4}$ years but that the luminosity will be driven by the Kelvin-Helmholtz collapse of the
extended envelope.  Their expectation was for the star to become significantly more
luminous ($500-5000 \;\Lsun$ for a MS 1 \msun{} star).  

Ideally, to study the long-term evolution of the impacted companion, one would want to model the three-dimensional interaction of the SN ejecta with a suite of
possible donor stars until the companion is relaxed and roughly spherical.  Then one would use the mean density, temperature, and composition
profiles as the initial conditions for a one-dimensional stellar evolution code and follow the companion until it settles back on the MS at the location of its
current mass.  \citet{ivanova04} show that the mass of the donor star can range from $\sim 0.6 \-- 1.5 \; \msun$.  We only explore the 1 \msun{} MS model of \citet{marietta00} because that paper contains the information needed to calibrate our model. Hydrodynamic simulations to calibrate our simple models at other masses or evolutionary states are not available, but the results presented in this paper should generally hold.  

We study this problem using the approach of \citet{podsiadlowski03}, who followed the evolution of a star which loses mass and is heated over a short period of time in order to mimic the effects of the explosion.
We used the stellar evolution code MESA (\citealt{paxton11}), allowing for fluid velocities.  We added a module to first 
have a rapid phase of mass loss followed by a rapid phase 
of extra heating. In the the heating phase we also included an 
outer boundary $R_{\textrm{out}}$, adjusting the mass loss rate so
that any material at $R>R_{\textrm{out}}$ is lost in a wind on
the escape time scale from that radius.  This allowed us
to treat mass loss as in \citet{podsiadlowski03} or to allow the heating phase to be responsible for some or all of the
mass loss when the 
amount of energy added becomes large.  Thus, the models
had 4 parameters: the amount of mass lost in the first
phase ($\Delta M$), the amount of energy added in the 
second phase ($E_{\textrm{heat}}$), the truncation radius ($R_{\textrm{out}}$),
and the time scale used for both the mass
loss and heating phases ($\Delta t$).  In each interval, the mass
loss and heating rates were linearly increased and
then symmetrically decreased. The internal energy added per unit
mass was distributed as the ratio of the enclosed mass
at any radius to the total mass.  Thus more energy is 
deposited at the edge of the star than near the center.
This leads to results more consistent with the \citet{marietta00} 
simulations than, for example, making the energy added per unit
mass independent of radius.  It also makes sense physically, as the shock will deposit more energy per unit mass in the lower density regions of the star. 

\begin{figure}[htp]
	\centerline{
		\includegraphics[height=8.0cm]{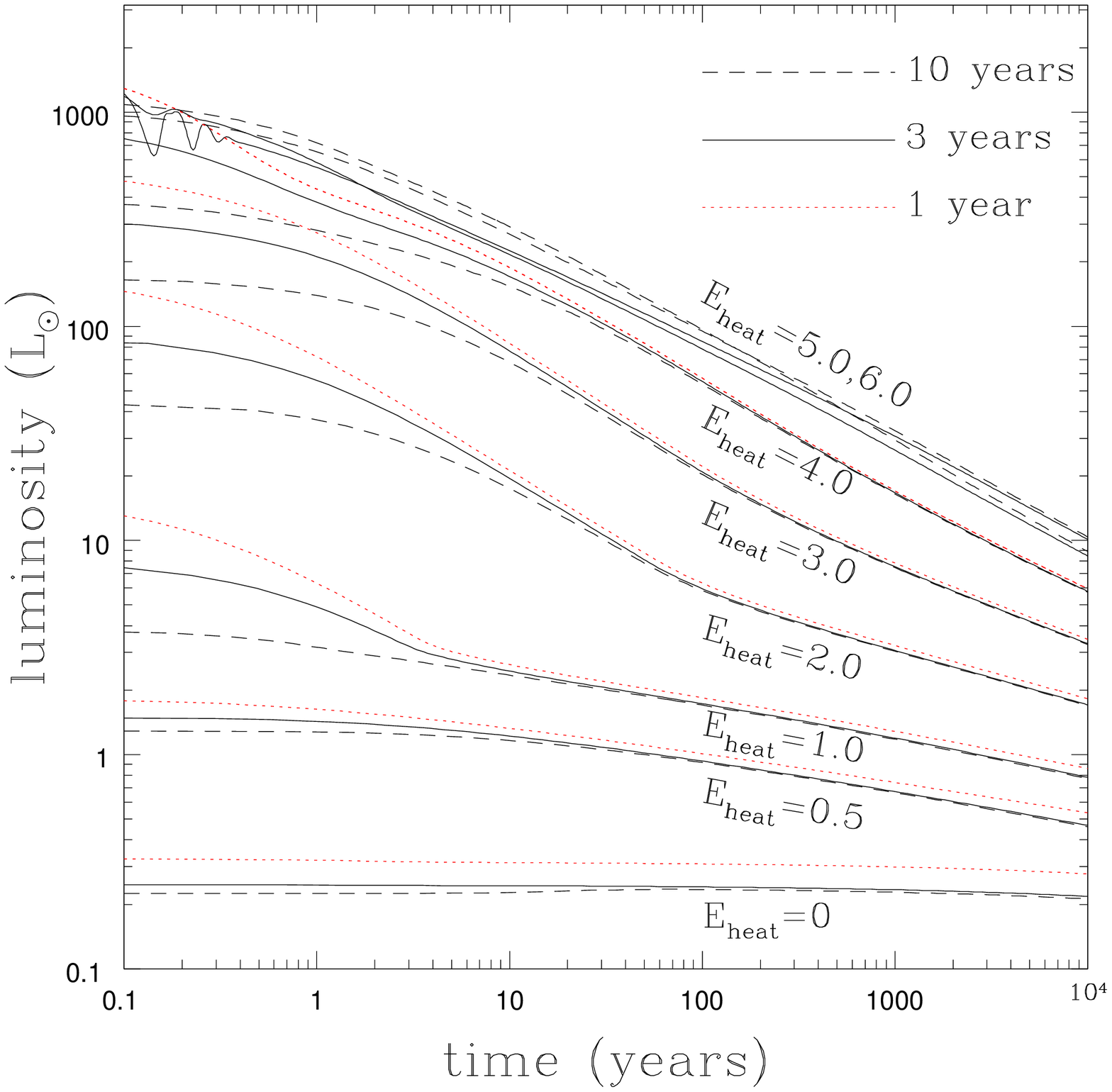}
	  }
	\centerline{	
		\includegraphics[height=8.0cm]{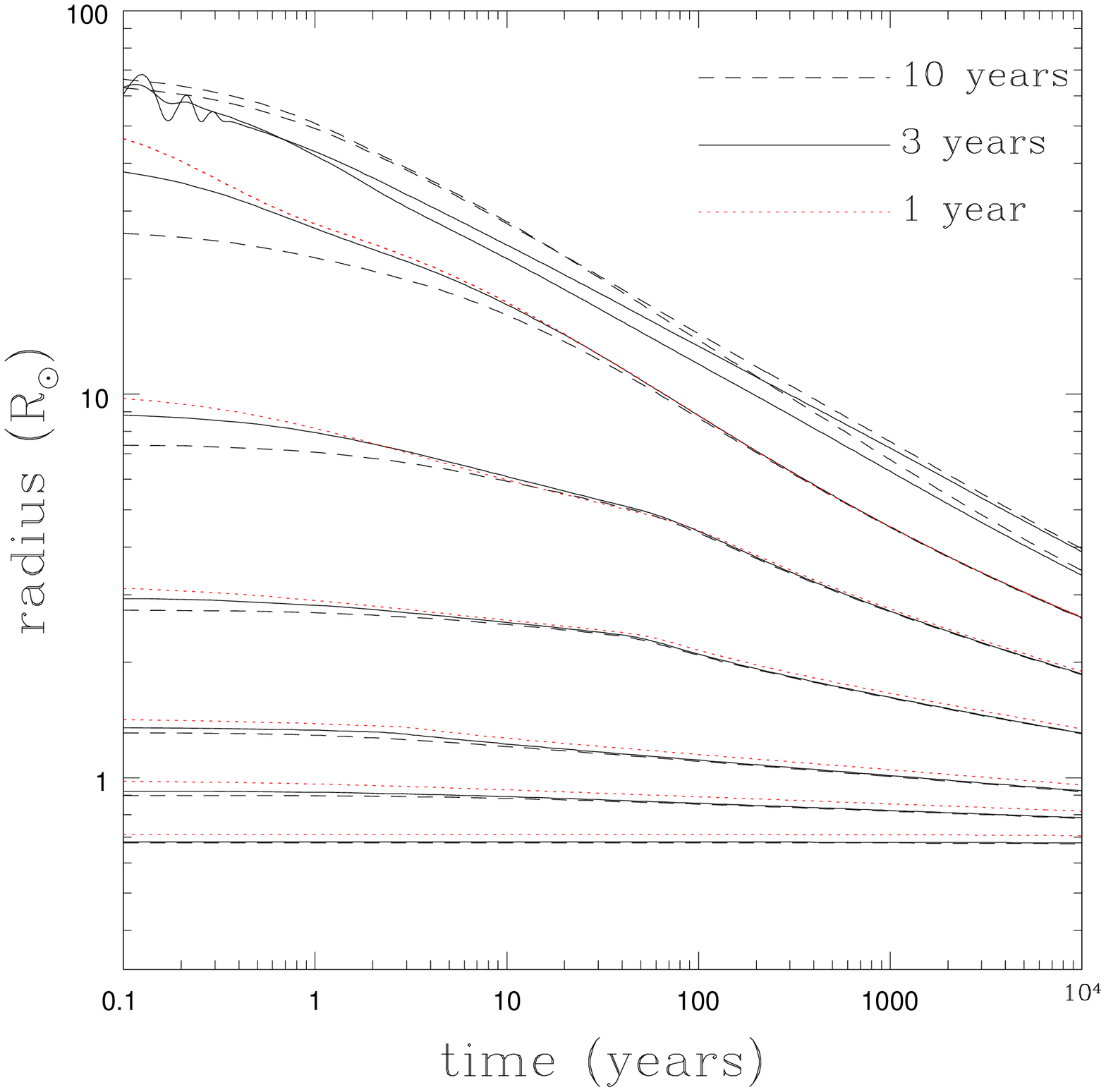}
	}
	\caption{The evolution of a 1 \msun{} MS companion after the stripping and heating phases have ended as a
	function of the energy added, $E_{\textrm{heat}}$ (in units of $10^{47}$~ergs), and 
	for a range of heating/stripping time scales $\Delta t= 1$, $3$ and $10$ years. These models have $R_{\textrm{out}}=100 \; \rsun{}$ and the
	radii of the highest $E_{\textrm{heat}}$ models expand to this limit, initiating a small amount of additional mass-loss and leading to very similar 
	subsequent evolution for these models.    {\it Top Panel:} Evolution of the companion's luminosity. For the same
	$E_{\textrm{heat}}$, the models converge after a time 
	$\Delta t$ has elapsed.  {\it Bottom Panel:} Evolution of the companion's radius.  The ordering of the cases is the same as in the top panel.  The peak
	luminosity is mostly determined by the maximum radius of the star.}

	\label{fig:LR}
\end{figure}

We first considered models where $\Delta M=0.15 \; \msun{}$ to match
the \citet{marietta00} hydrogen cataclysmic variable (HCV) model\footnote{
In the HCV model of \citet{marietta00} the binary separation was $3$ \rsun{}, the secondary was represented by a $1$ \msun{} main sequence model, and the SN  was represented by the W7 ejecta profile \citep{nomoto84}.}
 and $E_{\textrm{heat}}=0$, $0.5$, $1$, $2$, $3$, $4$, $5$ or $6 \times 10^{47}$~ergs.  We used time scales of $\Delta t=1$, $3$ and
$10$ years, and  $R_{\textrm{out}}=10$, $30$ and $100 \; \rsun{}$.  The $E_{\textrm{heat}}=5$ and $6\times 10^{47}$~ergs models did not converge
for $\Delta t=1$~year and $R_{\textrm{out}}=100 \; \rsun{}$  due to the 
development of an oscillation as the transient ends and the
outer envelope collapses.  Figure \ref{fig:LR} shows the evolution of the companion in luminosity and radius as a function of $E_{\textrm{heat}}$ and the
duration $\Delta t$ assuming $R_{\textrm{out}}=100 \; \rsun{}$ after the transient phase ended.  The expected luminosity and radius evolution
depends little on $\Delta t$ once a time period $\Delta t$
has elapsed after the transient phase.  The main effect appears
to be an underestimate of the luminosity in that period, 
presumably because of the additional energy radiated during
the transient phase.  Starting with the $4 \times 10^{47}$~erg
model, the star expands to $R_{\textrm{out}}=100 \; \rsun{}$, leading to a
small amount of additional mass loss.  The inclusion of the
outer boundary appears to drive the high $E_{\textrm{heat}}$ models to
a similar final structure so that there is little difference
in their subsequent evolution.  
The luminosity of the higher $E_{\textrm{heat}}$ models can
be restricted by reducing $R_{\textrm{out}}$ because it is the rapid
collapse of a greatly over-expanded envelope that produces the
highest luminosities.  For example, using $R_{\textrm{out}}=10 \; \rsun{}$
caps the luminosity of the high $E_{\textrm{heat}}$ models at 
roughly $80 \; L_\odot$, but they sustain that luminosity
for a long time and then roughly merge onto the evolution
of the models with $R_{out}=100 \; \rsun{}$. 

\begin{figure}[htp]
	\includegraphics[height=16cm]{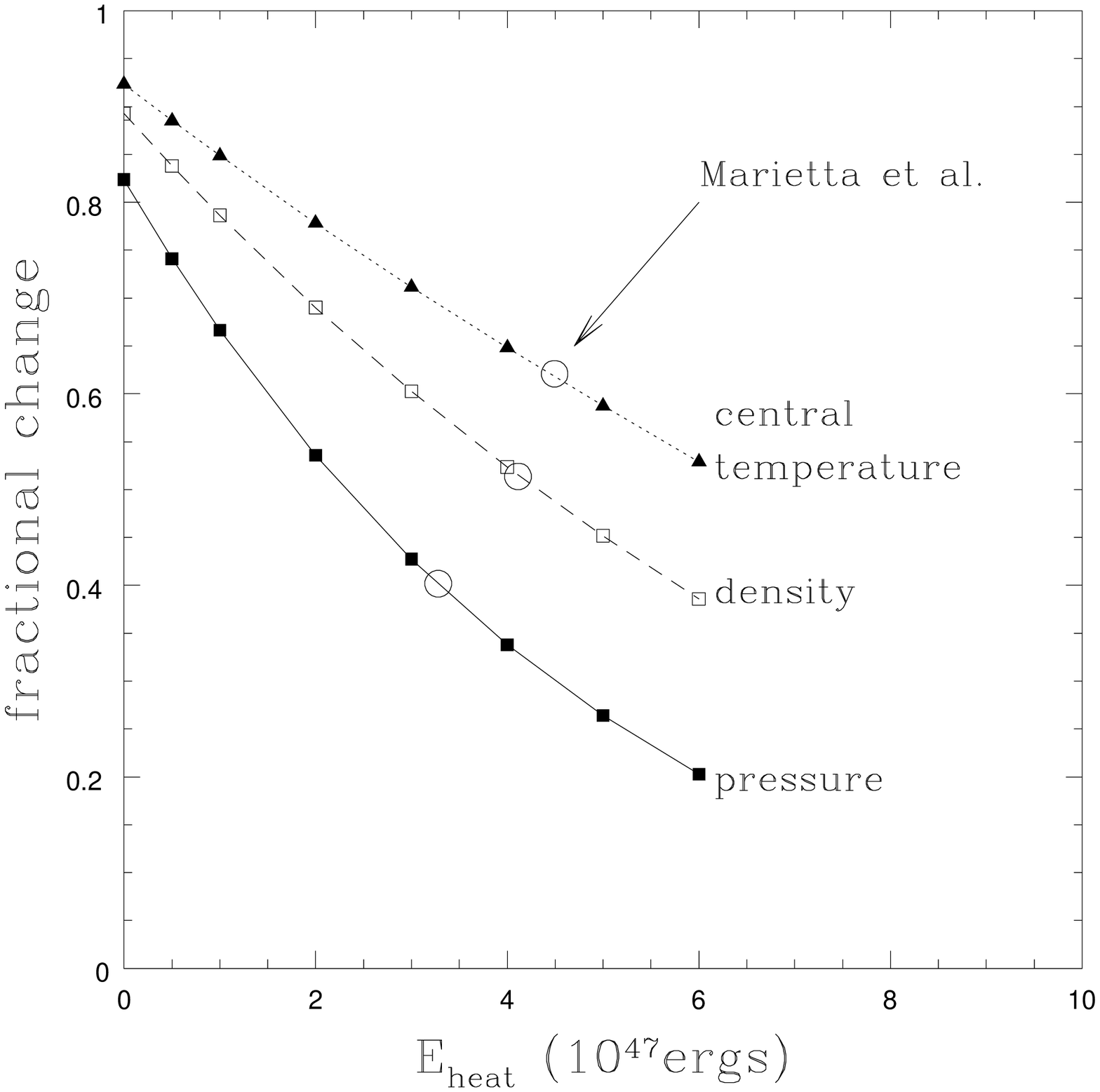}

	\caption{The fractional change in the central temperature (triangles, dotted),
	density (open squares, dashed), and pressure (filled squares, solid) of the
	model as a function of the energy added $E_{\textrm{heat}}$.  The large circles on
	each curve mark the fractional change in the HCV model of \citet{marietta00},
	which is generally achieved for $E_{\textrm{heat}} \simeq 4 \times 10^{47}$~ergs.  
	Here we show the results for the $\Delta t =3$~years, $R_{\textrm{out}}=100 \; \rsun{}$ models, although
	the results depend little on these parameters.  }
	\label{fig:central}
\end{figure}

\begin{figure}[htp]
	\includegraphics[height=16cm, angle=90]{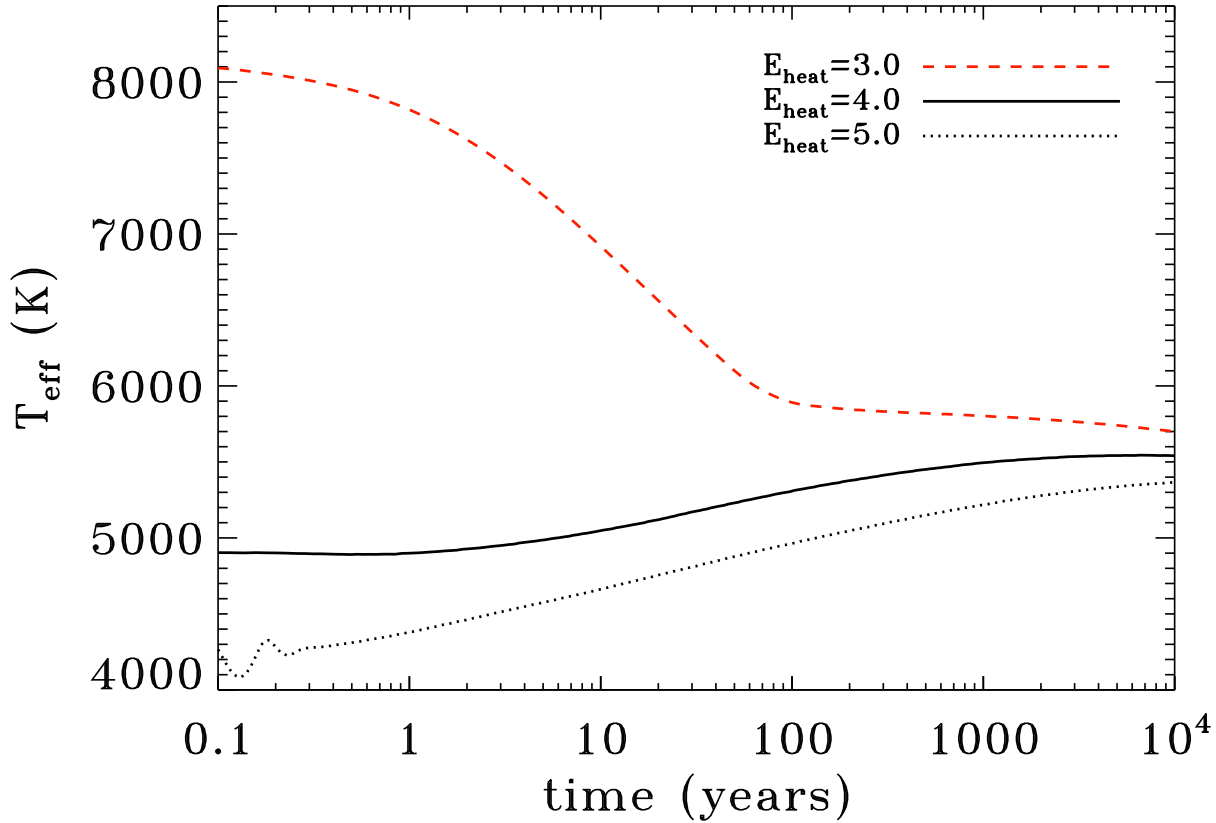}

	\caption{Similar to Figure \ref{fig:LR} except showing the evolution of the companion star's effective temperature for three $\Delta t=3$ years models. $E_{\textrm{heat}}$ is in units of $10^{47}$~ergs. Note that for larger (smaller) amounts of energy injected into the companion's envelope, the star becomes cooler (hotter).}
	\label{fig:T}
\end{figure}

Figure \ref{fig:central} presents the fractional change in the central temperature, density, and pressure for the $\Delta t=3$~year, $R_{\textrm{out}}=100 \; \rsun{}$
models as a function of the energy added, $E_{\textrm{heat}}$.  This figure shows that $E_{\textrm{heat}} \simeq 4 \times 10^{47}$~ergs is needed to reproduce the fractional change in the HCV model of \citet{marietta00}.  For the remainder of the paper we will take the $E_{\textrm{heat}} = 4 \times 10^{47}$~ergs, $\Delta t=3$~year, and $R_{\textrm{out}}=100 \; \rsun{}$ model as our fiducial companion model.  Depositing the energy per unit mass uniformly or significantly changing $E_{\textrm{heat}}$ from this value, leads to models that do not reproduce the \citet{marietta00} simulations well.  Figure \ref{fig:T} 
shows the evolution of the effective temperature ($T_{\textrm{eff}}$) for the fiducial model and models with higher and lower $E_{\textrm{heat}}$.  In general, we find that if $E_{\textrm{heat}}$ is large enough to initiate mass loss (the fiducial and higher $E_{\textrm{heat}}$ models) then the temperature decreases, and models with insufficient heating to initiate mass loss (the lower $E_{\textrm{heat}}$ models) become hotter.  Since the ultimate mechanism of the mass loss is ablation due to heating, only the models with enough energy input to initiate mass loss are realistic.  The total energy in SNe Ia ejecta is $\sim 1.2 \times 10^{51}$~ergs, 
of which $\sim 3.3 \times 10^{49}$~ergs is incident on the star since, assuming the companion must overflow its Roche-lobe, $a/R \simeq 3$.  The binding energy of the $0.15$ \msun{} of stripped material is $E_{\textrm{strip}} \simeq 6 \times 10^{47}$~ergs, so the total energy transfered to the star is approximately  $E_{\textrm{heat}}+E_{\textrm{strip}} \simeq 10^{48}$~ergs and $\simeq 60\%$ of that energy is used to strip material
from the surface, and $\simeq 40\%$ heats the remaining star. Only a small fraction ($f \simeq 0.03$) of the incident shock energy is transferred
to the star. Simply assuming that $3\%$ of the shock energy is transfered to the star with $60\%$ of that energy going into stripping mass does not perfectly reproduce the scaling of
mass loss with $a/R$ in \citet{marietta00}, but it comes considerably
closer to doing so than the analytic estimates of \citet{wheeler75}. 
More generally, the transfer of energy into heating the surviving
star is a crucial parameter that needs to be reported for any
future simulations.
  
In the final analysis, the basic physics is relatively easy to
understand.  The shocks heat the star, depositing more energy
per unit mass near the surface than near the center.  However,
the shocks have no physical knowledge of which material will
eventually escape and the star has no sharp density jumps to
create any features in the energy deposition.  The heated material then
expands, but with a continuous density distribution between material which
ultimately escapes and material which remains bound.  As a result,
the energy deposited in material which eventually escapes 
cannot be significantly larger than the energy deposited in the
material which does not escape. Mass
is stripped beyond some effective radius $R_{\textrm{out}}$, leaving
a puffed-up envelope that then begins to cool and collapse.
The more extended this envelope, the higher the peak luminosity
since a lower density envelope has a shorter thermal time scale.
Ultimately, however, for the same envelope extent the star
produces similar luminosities.  Because there are no means
in this process of distinguishing between escaping and bound
material until the escaping material is stripped, the low
$E_{\textrm{heat}}$ models following \citet{podsiadlowski03} are 
unphysical because they effectively make the heating of the
unbound material independent of the heating of the bound 
material.  When the mass loss is driven by ablation, the energy used to strip mass must be tightly correlated with the energy used to heat the surviving star and is not an independent variable. 

\begin{figure}[htp]
	\includegraphics[height=16cm]{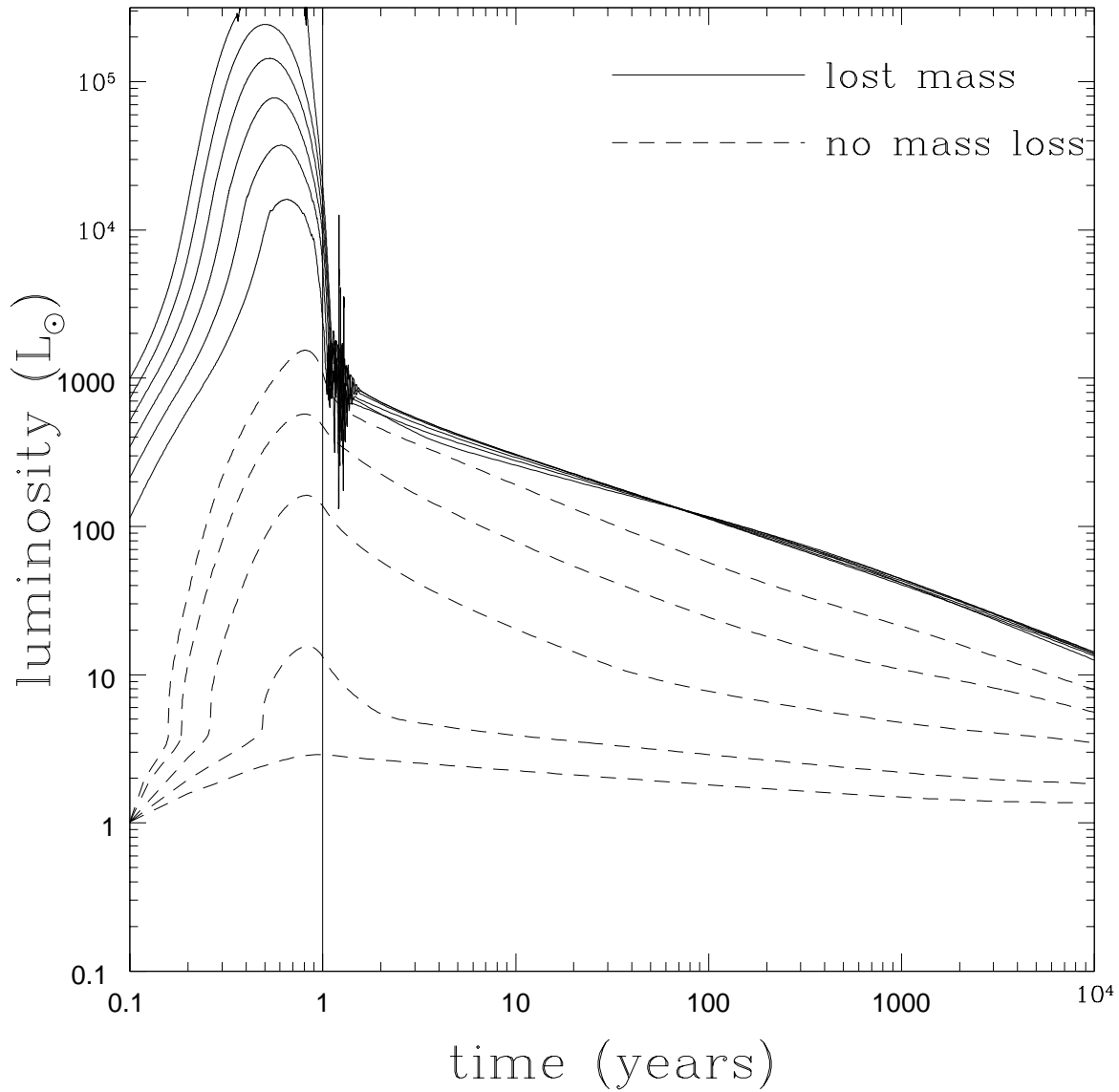}

	\caption{The evolution of the luminosity for models with increasing 
	values of $E_{\textrm{heat}}$ where the mass loss is drive only by the heating.
	The models shown by the solid lines have lost mass, while those shown
	by the dotted lines have not expanded enough to lose any mass.  Note
	how all models with mass loss show similar late time evolution.  The
	values of $E_{\textrm{heat}}$ are, from bottom to top, $0.5$, and $1,2,\ldots,10 \times 10^{47}$~ergs,
	$R_{\textrm{out}}=100 \; \rsun{}$ and $\Delta t=1$~year.  The vertical line marks
	the end of the heating period.  The rapid collapse of the envelope
	leads to some initial oscillations that do not seem to affect the
	general results.}
	\label{fig:Rout}
\end{figure}

In this sense, alternate models where we drive the mass 
loss with only heating are far more physical.    For small
values of $E_{\textrm{heat}}$ the star expands but does not reach
$R_{\textrm{out}}$ and has no mass loss.  Larger values of $E_{\textrm{heat}}$ ($> 4 \times 10^{47}$ ergs)
expand the star to $R_{\textrm{out}}$, and the stars begin to 
lose mass.  The post-transient evolution of {\it all}
stars that have had mass loss is very similar because 
the primary driver of their luminosity
is the collapse of the envelope from $R_{\textrm{out}}$ to its
eventual equilibrium radius.  This is illustrated in
Figure \ref{fig:Rout}, which shows the evolution as a function of
$E_{\textrm{heat}}$ labeled by whether or not the stars lost
any mass assuming $R_{\textrm{out}}=100 \; \rsun{}$.  Ignoring the 
luminosity during the transient, much of which is 
radiating energy injected into the expanding envelope,
the subsequent evolution of the mass losing transients is essentially
independent of the deposited energy.  The amount of mass
lost rises steadily with $E_{\textrm{heat}}$, and for the maximum
energy $E_{\textrm{heat}}=10^{48}$~ergs, the mass loss is approximately
the same as in the HCV model.  The central pressure of this
model is, however, too low, which means that we should
concentrate more of the energy deposition towards the 
surface. 

\section{Implications for Nearby Type Ia Supernovae}
\label{sec:Individual}

Although challenging, searching for the donor star after a \snia{}
explosion has significantly faded is a popular method for testing the SD
model. This can be done either by examining sources near the center of historical SNRs or by observing the sites of recent \sneia{} in external galaxies. What has not been clearly factored into these discussions is that the surviving star must be cooler and more luminous than it was before the explosion.  This is an unavoidable consequence of mass loss by ablation.  Our fiducial model with $E_{\textrm{heat}} \simeq 4 \times 10^{47}$ ergs and $R_{\textrm{out}} = 100 \; \rsun{}$, the best match to simulations (shown in Figures \ref{fig:LR} and \ref{fig:T}), indicates that initially after the SN the companion has $L \sim 300 \-- 1000 \;\Lsun{}$ and $T_{\textrm{eff}} \sim 5000-6000$ K, depending slightly on $\Delta t$.  Then, after $\sim 10$ years, we find $L \sim 100 \-- 200 \;\Lsun{}$ and $T_{\textrm{eff}} \sim 5000 \-- 5200$ K. After $\sim 100$ years, we see $L \sim 50 \-- 60 \;\Lsun{}$ and $T_{\textrm{eff}} \sim 5300 \-- 5400$ K. Finally, after $\sim 1000$ years $L \sim 15 \-- 20 \;\Lsun{}$ and $T_{\textrm{eff}} \sim 5500$ K.

\subsection{Supernova Remnants}

Searching for surviving stars near the centers of SNRs is a promising method to distinguish
between progenitor models \citep{canal01, ruiz-lapuente04}, particularly given the ability to determine the SN type using either X-ray spectra of the remnant \citep{hughes95, badenes06, badenes08} or dust echo spectra of the actual event (e.g., \citealp{rest08a}).  Given \snia{} rates and SNR lifetimes, there are several examples in both the Galaxy and the Magellanic Clouds. The challenge is to precisely identify the expected location of the star. First, the companion will have a non-negligible
proper motion, which can be in any direction outwards from the SN
explosion site \citep{marietta00}.  Second, while in general Type Ia SNRs are strikingly spherical,
they will be slightly asymmetric due to differing densities of the
surrounding interstellar medium (ISM) and/or
bubbles or clumps of gas and dust in the surrounding environment, which will add uncertainty to the
measurement of the geometric center of the SNR.  Moreover, as time passes, the donor star fades as it returns to thermal equilibrium.

\subsubsection{SN 1572: Tycho's Supernova}
\label{sec:Tycho}

SN 1572, first seen on November 6th, 1572, was well observed for two years (e.g., Tycho Brahe, see \citealp{ruiz-lapuente04b}).  Both the light curve
\citep{ruiz-lapuente04b} and the spectrum obtained from light echoes
\citep{rest08b, krause08} show SN 1572 to have been a normal \sneia{}.  In the past decade there have been many attempts to identify a SD
companion star to SN 1572.  \citet{ruiz-lapuente04} performed the
first search and found a promising candidate near the center of Tycho's SNR, Tycho G, a sub-giant with spectral
type G2 IV, which has $T_{\textrm{eff}}=5750 \pm 250$ K, $M \approx 1 \; \msun{}$, $R \approx 1 \-- 3 \;\rsun{}$, and surface gravity log($g$/cm s$^2$)$\sim 3.0 \-- 4.0$.  They found that it is at a
distance of $d \approx 2.5-4.0$ kpc, with a large radial velocity of
$v_{\textrm{r}} = -108 \pm 6$ km s$^{-1}$, and a high derived
proper motion of $94 \pm 27$ km s$^{-1}$.
However, there is significant controversy about the
true nature of Tycho G.  \citet{fuhrmann05} proposes that Tycho G
might simply be a thick-disk star coincidentally passing in the
vicinity of the SNR.  \citet{ihara07} classify Tycho G as a F8 V star
and argue that it could not be the donor star of SN 1572 because it
does not exhibit the blue-shifted Fe I absorption lines
expected for a star surrounded by the ejecta of a SN due to Fe I in the ejecta.  The lack of Fe I absorption suggests that Tycho G is a
foreground star. \citet{kerzendorf09} measure a smaller radial velocity ($79 \pm 2$ km
s$^{-1}$) and detect no proper motion, which is in moderate disagreement with the
\citet{ruiz-lapuente04}, and suggests that Tycho G is not
associated with the SN event. \citet{gonzalezhernandez09} used
Keck high-resolution optical spectra to improve the stellar
parameters, finding $T_{\textrm{eff}}=5900 \pm 100$ K, log($g$/cms$^2$) $=
3.85 \pm 0.30$ dex, and $v_{\textrm{r}} \sim 80$ km
s$^{-1}$. They also find that Tycho G is overabundant in Ni, suggesting that it was polluted
by the SN event.  

The rotational velocity ($v_{\textrm{rot}}$) of a surviving donor star could also be
an important diagnostic of \sneia{} progenitor systems in the SD model.  \citet{meng11} predict that surviving companions should be rapidly rotating with $v_{\textrm{rot}} \sim 100$ km s$^{-1}$ because they are tidally locked at the time of the explosion.  Tycho G, however, rotates slowly with estimates of  $v_{\textrm{rot}} = 7.5 \pm 2$ km s$^{-1}$ and
$v_{\textrm{rot}} < 6.6$ km s$^{-1}$ by \citet{kerzendorf09} and
\citet{gonzalezhernandez09}, respectively. \citet{meng11} did not
account for the interaction of the SN ejecta with the donor, and
they suggest this might explain the difference. In their full simulation with rotation and orbital motion, \citet{pan12} found that the impact removes $48 \%$ of the angular momentum from a MS companion but only $18 \%$ of its mass.  They estimate an equilibrium radius for the donor star after the SN of $R \simeq 2.4 \;\rsun{}$, which still implies a high rotational velocity of $v_{\textrm{rot}} \sim 37$ km s$^{-1}$.
They suggest that the donor will still be out of
thermal equilibrium and thus have a larger radius. However, since the angular
momentum scales with radius and rotational velocity as $J \propto M \ R \ v_{\textrm{rot}}$, the stellar radius must be $R \ge 9 \; \rsun{}$ to be consistent with the observed limits.  This conservatively assumes solid body
rotation and a self-similar radial distribution of mass in
the donor.  However, $R \simeq 9 \;\rsun$ is completely ruled out by the observed luminosity and temperature, so there remains a large disagreement between the observed and expected $v_{\textrm{rot}}$ of Tycho G assuming it was the donor star in the SN 1572 progenitor system.

To this list of problems we can now add that Tycho G is too low luminosity and too high temperature to be the survivor of a Ia explosion.  For any of our models consistent with \citet{marietta00}, the luminosity should be $\ge 20 \; \Lsun{}$, rather then the estimated $2 \-- 8 \; \Lsun{}$, and the temperature should be $\le 5500$ K, rather than the observed $5900 \pm 100$ K. The only models which would agree with the observed values are essentially the ``no heat'' models that are grossly inconsistent with the initial conditions demanded by the simulations.  Based on its luminosity, temperature and rotation rate, we conclude that Tycho G is not associated with SN 1572 and that SN 1572 was not a traditional SD SN.

\subsubsection{SNR $0509-67.5$}
\label{sec:LMC1}

In many ways, the Large Magellanic Cloud (LMC), at a distance of $50.6 \pm 1.6$ kpc \citep{bonanos11}, is a better place to look for the companions to Type
Ia SNRs because the distances to the remnant and nearby stars are well-determined. Moreover, it is still close enough that Hubble Space Telescope ({\em HST}\/) or ground-based adaptive optics
observations can search for lower mass main sequence stars without severe problems from crowding. 

The central region of the LMC  SNR $0509-67.5$, created by a 1991T-like
\sneia{} explosion $400 \pm 50$ years ago \citep{hughes95,
rest05, rest08a, badenes09}, was recently studied by
\citet{schaefer12} using archival $B$, $V$, $I$, and H$\alpha$  {\em HST}\/
images.  They were able to constrain the search region to $1 \farcs 4$ and found no ex-companion stars in the central
region to a limiting visual magnitude of $M_{V} = +8.4$ mag, which they
claim rules out all published SD models. However, this limit does not to account for the line of sight extinction to the LMC ($A_{V} \approx 0.25$ mag), so the actual limit is $M_{V} = +8.15$ mag \citep{distefano12}.  Additionally, \citet{schaefer12} assumed that main sequence donor stars will be minimally affected by the explosion.  As with SN 1572, our fiducial models calibrated to agree with \citet{marietta00} predict that the surviving star should be $\sim 20 \; \Lsun{}$ after $\sim 500$ years.  Crudely speaking, this means that a limit of $M_{V} > +8.15$ mag now corresponds to a limit of $M_{V} \gtrsim +11.25$ on the unperturbed progenitor star, which would correspond to a $M \simeq 0.25 \; \msun$ dwarf \citep{marigo08}.   Lower mass companions are both rare and will be more affected by shock heating, so our conclusions largely hold for their scenarios as well.

\subsubsection{SNR $0519-69.0$} 
\label{sec:LMC2}

The central region of the LMC SNR $0519-69.0$ was studied recently by
\citet{edwards12} using archival $V$ and H$\alpha$  {\em HST}\/ images. Light
echoes of SNR $0519-69.0$ show that the SN spectrum is that of a
normal \sneia{} that exploded $600 \pm 200$ years ago \citep{rest05, edwards12}.
Unfortunately, \citet{edwards12} were only able to constrain the
location of possible companions to within $4 \farcs 7$.  In this region, \citet{edwards12} found 27 MS stars brighter then $V = 22.7$ mag.  The brightest star, Star 1, with $V = 19.7$ mag corresponds to  $M_{V} > +1.2$, which would allow for a solar mass companion. However, adding the expectation that the luminosity must be roughly increased by a factor of 10, this really corresponds to a limit on the progenitor of  $M_{V} \gtrsim +4$ and a solar mass companion is only marginally allowed. The limits are much more stringent for the fainter stars in the error circle. Additionally, with a $V - H_{\alpha}$ color of $\simeq 0$, Figure 2 of \citet{edwards12} shows that this star is on or close to the main sequence.  Based on our fiducial model we would expect a ex-companion to be redward of the main sequence, suggesting that Star 1 is not the associated with the SNR.  However, further studies of the possible companions with more {\em HST}\/ observations at different wavelengths or spectra should be carried out.  Currently, the nature of the progenitor system of SNR $0519-69.0$ can be still be considered uncertain.

\subsection{Recent Supernovae}

Besides observing the explosion sites of historical SNRs, one could also look at the location of more recent \sneia{}
once the SN has faded.  This method benefits both from having very accurate search positions and from the greater luminosity of the surviving stars at earlier times ($L \sim 50 \-- 300 \;\Lsun{}$).  The main problem is that these directly observed SN are far more distant, with the closest being at a distance of $> 3$ Mpc.  We briefly discuss four of the closest \sneia{}
observed in the last century.

\subsubsection{SN 1937C, SN 1972E, and SN 1986G}
\label{sec:SNe}

Three historic \sneia{}, SN 1937C, SN 1972E, and SN 1986G, are particularly good candidates for a study of their post-explosion systems because they are nearby and were well observed. SN 1937C in IC 4182 at $\sim 4.4$ Mpc \citep{saha06} was discovered by Zwicky in the fall of 1937 at Palomar Observatory \citep{baade38}.  SN 1972E was discovered in NGC 5253 at $\sim 4.2$ Mpc \citep{saha06}.  Finally, the closest \snia{} discovered in the digital imaging era, SN 1986G, was discovered in NGC 5128 (Centaurus A) at a distance of $\sim 3.4$ Mpc \citep{evans86, ferrarese07}.  Given the historical SN explosion images, it should be possible to centroid the SN explosion to a fraction of an arcsecond and with {\em HST} observations one could plausibly find or rule out a SD companion for these \sneia{}. 

\subsubsection{SN 2011fe}
\label{sec:2011fe}

The recent Type Ia SN 2011fe in M~101 is the nearest Ia in the last 25
years. At a mere 6.4 Mpc \citep{shappee11}, this SN Ia provides an
exceptional opportunity to put the strongest constraints on the
progenitor system of any SNe Ia. \citet{patat11}
obtained 12 epochs of high resolution spectra of SN 2011fe, finding that 
it is only slightly reddened and lies in a
``clean'' environment.  \citet{horesh12} used upper limits from both
radio and X-ray observations of SN 2011fe to rule out circumstellar
material from a giant companion. \citet{chomiuk12} then strengthened
these radio constraints with EVLA observations, ruling out most
popular SD progenitor models.  The proximity and the low extinction of
SN 2011fe allowed \citet{li11} to place $10\--100$ times stronger
constraints on the visible light from the progenitor system than previous
studies using deep archival pre-explosion {\em HST}\//ACS
observations.  These observations rule out luminous red giants ($ >
3.5 \; \msun$) and most helium stars as plausible donors.

SN 2011fe was discovered very quickly by the Palomar Transient Factory
\citep{law09} on August 24, 2011, less than one day after explosion, allowing multi-wavelength follow-up observations to be quickly carried out.
With the early time UV, optical, and X-ray observations \citet{nugent11} and \citet{bloom12} concluded that the radius of the primary was smaller than $0.1 \;\rsun$ and $0.02 \;\rsun$,
respectively. \citet{bloom12} then went onto show that their
constraints leave compact objects as the only viable primary.
\citet{brown12b} and \citet{bloom12} also used the very early-time UV
and optical observations to constrain the shock heating which would
occur as the SN ejecta collides with the companion.  They rule out
giants and main sequence secondaries more massive than the sun.  These
constraints seem to imply that the DD model is the only viable
progenitor model.  However, these studies both rely on
the calculations of \citet{kasen10}, which assumes a Roche-lobe-overflowing secondary
with a typical stellar structure.  As previously discussed,
\citet{meng07} argued that the secondary should be smaller and more
compact than a typical main sequence star.  Such a smaller, more tightly
bound companion would produce a weaker signal at even earlier
times which could have been missed.  Additionally, slightly less massive companions can also be too small for the SN ejecta shock to have been observed.

\begin{figure}[htp]
	\includegraphics[height=16cm, angle=90]{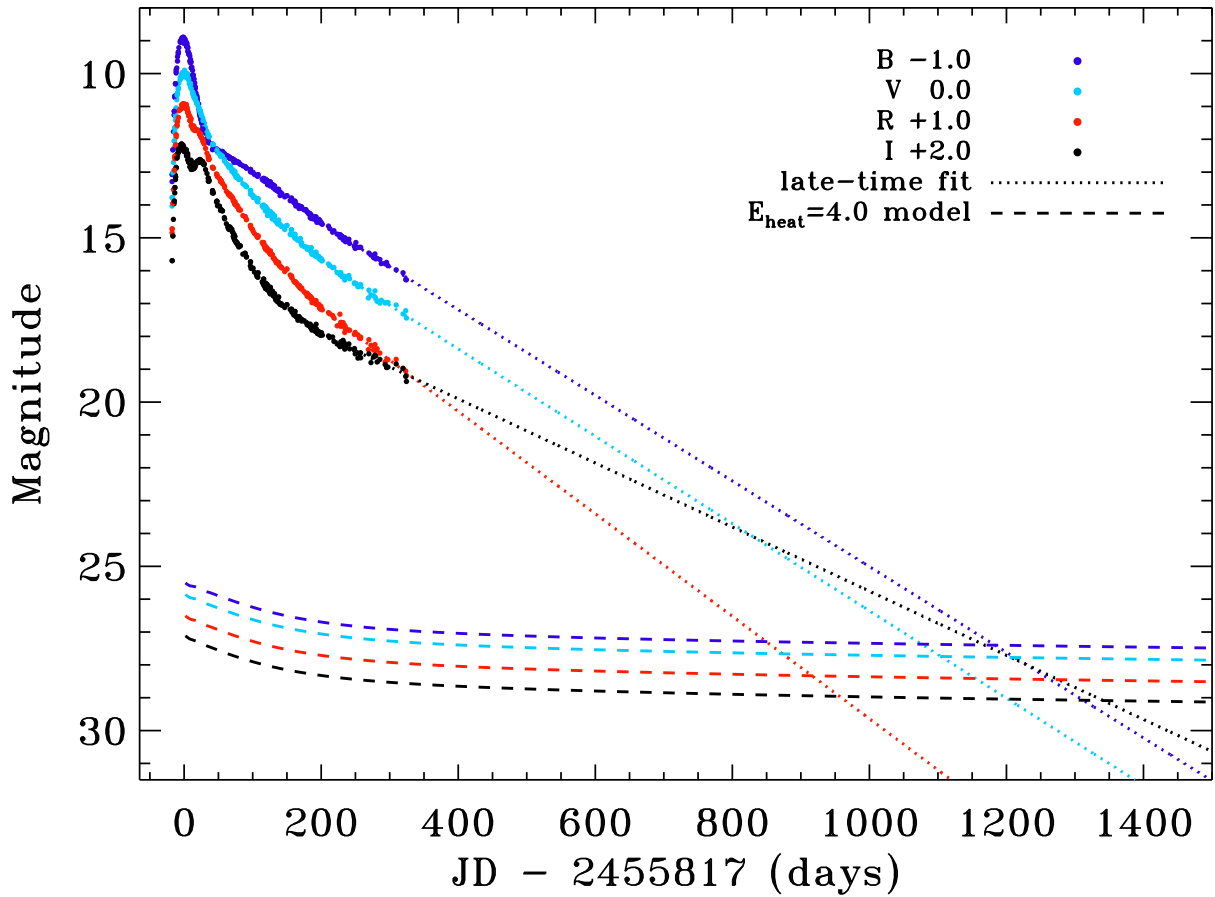}

	\caption{SN 2011fe light curve (points) with linear late-time decline fit (dotted black) and our fiducial model for a shock heated companion (dashed, see Section \ref{sec:PostProp}). We note that a signature of the presence of a SD MS companion to a \sneia{} explosion would be a break in the very late-time light curve ($\sim 900$ days) as the shock-heated companion becomes more luminous than the rapidly fading SN ejecta. We have assumed a blackbody to convert the models from \S\ref{sec:PostProp} into Vega magnitudes. }
	\label{fig:SNfe}
\end{figure}

Very late time observations of SN 2011fe would place further
constraints on the progenitor system by detecting, or putting upper
limits on, the existence of a shocked donor.  The luminosity of
the surviving donor star is expected to be $L\sim 300 \;\Lsun{}$, which corresponds to
$V \sim 27$ mag.  In Figure \ref{fig:SNfe}, we present the $BVRI$ light light curves of SN 2011fe of \citet{richmond12}, \citet{munari12}, and \citet{shappee12c} along with a linear fit for the late-time decline
as compared to the expected evolution of our fiducial model. If the current decline rate continues, the
shocked companion should begin to dominate the SN light in the $R \sim 90
0$ days after maximum light. At that point, {\em HST}\/ observations monitoring the very
late-time light curve of SN 2011fe should see it level off as most of
the observed flux will then be coming from the shock heated companion and not the
expanding, cooling shell of ejecta.  

However, there are other effects which might also cause the late-time SN light curve to flatten.  First, light echoes from the SN explosion off nearby dust can add additional late-time flux as was seen from SN 1987A \citep{crotts88}.  Fortunately, the environment surround SN 2011fe shows no evidence for gas or dust \citep{patat11}.  Second, Auger and internal conversion electrons emitted in the decay of $^{57}$Co are expected to become a significant source of heating $\sim 750$ days after the explosion and dominate the positron decay of $^{56}$Co after $\sim 1000$ days \citep{seitenzahl09, seitenzahl11}. The late-time bolometric light curve is then expected to flatten (Figure 4 of \citealp{roepke12}).  Fortuitously, the evolution of a nearby companion will occur on time-scales much longer than the half-life of $^{57}$Co (271.79 days). Thus, a single degenerate companion would still be observable at even later times.

\section{Core-Collapse Supernovae}
\label{sec:core}

Interestingly, a similar process may also be possible during stripped core-collapse supernova (SN IIb, Ib and Ic) explosions.  One of the theoretical models for producing stripped core-collapse supernovae progenitors is using binary mass transfer to remove the outer envelope of the primary (e.g., \citealp{yoon10}), a common envelope phase (e.g., \citealp{nomoto95}), or both. In such scenarios, the companion is again forced to lie relatively close to the primary, with similar issues about mass stripping and shock heating to the SD \sneia{} case. 

The maximum shock energy is essentially the same as for the Ia case because it is determined by the geometry of Roche-lobe overflow.  This means that the effects on main sequence binary companions, are likely greatly reduced.  First, the
escape velocity of a higher mass star is somewhat larger, so
less mass is stripped for a given amount of energy.  Second, the
internal energy of the higher mass star is much larger, so the 
fractional change in the energy created by the shock is much 
smaller.  Third, the luminosity of the star is much larger, so
any adjustment leading to a roughly constant source of additional
luminosity is a much smaller fractional change in the luminosity.
As a result, we would expect that the effects on an $8 \; \msun{}$ main sequence companion
with the same shock energy and geometry will not very dramatic, and this is confirmed in our numerical experiments.  Additionally, the typical ccSN binary companion (without mass transfer) can be at much larger distances than in a \sneia{}, so the effects will likely be
unobservable for MS companions.  

However, if the companion has evolved into a giant star, the effects can be much greater due to the significantly reduced binding energy of the companion's envelope. It may even be possible, if the separation between the SN and a giant companion is small, for the SN ejecta to completely strip the envelope of the companion, leaving only the dense core, as seen in hydrodynamic simulations of \sneia{} explosions with a giant companion (e.g., \citealp{marietta00}).  The problem is that most of the time the companion star to a ccSN will still be on the main sequence (see \citealp{kochanek09}) unless the binary members were nearly exact twins \citep{pinsonneault06}.

\section{Conclusion}
\label{sec:conclusion}

Main-sequence (MS) donor stars surviving the explosion of a SD \sneia{} must be strongly heated and puffed up in a manner tightly coupled to their mass loss. In stellar evolution models that roughly match the final state of a 1 \msun{} star losing 0.15 \msun{} in the hydrodynamic simulations of \citet{marietta00}, the star must be heated by $E_{\textrm{heat}} \simeq 4 \times 10^{47}$~ergs, and it commences its post explosion evolution with the radius and luminosity dramatically increased and the effective temperature modestly decreased. When mass is lost by ablation, it is not physically possible to avoid these effects.  This means that the ``leftover'' stars of Tycho's SN and in the LMC supernova remnants SNR $0609-67.5$ and SNR $0519-69.0$ should still be $\sim 20$ times more luminous than before the explosion.  This greatly strengthens the arguments of \citet{schaefer12} against a SD scenario for the LMC SNR $0609-67.5$ and essentially rules out Tycho Star G as the survivor of SN 1572. Finally, such luminous donors should be observable with {\em HST}\/ within $\sim 10\;$Mpc.  We point out that there are a number of recent nearby Type Ia SNe, including SN 1937C, SN 1972E, SN 1986G, and SN 2011fe $\sim 2$ years after the explosion that should be studied in detail. For the latter, difference imaging between late time and pre-explosion {\em HST}\/ images should be a particularly powerful probe.  Thus, the SD channel has been ruled out for at least two nearby \sneia{} far more strongly than assumed and can be easily tested for several more. 

Recent works have suggested the ``spin-down model'' as a way to avoid the existing limits by using rotational support of the WD to delay the explosion long enough to render the companion far fainter than at the time of accretion \citep{justham11, distefano11, distefano12}.  In addition to several problems of fine tuning, there are no known physical systems where accretion leads to rotational velocities sufficiently close to breakup for rotation to be of much relevance to the hydrostatic support of the system.  The fastest spinning WD known (RX J0648.0--4418, $P_{\textrm{spin}} \simeq 13.2$ s, \citealp{mereghetti11}) is still spinning at less than half the rate ($P_{\textrm{spin}} \lesssim 6$ s) needed for rotation to be important \citep{yoon05}.  Moreover, most of the scenarios considered by \citet{distefano12} lead to lower mass companions which will be even more drastically affected by shock heating, so our conclusions largely hold for their scenarios as well.

Existing hydrodynamic simulations of the effects of explosions on \sneia{} companions have largely ignored the consequences of the explosions for the future evolution of the stars, although \citet{marietta00} comment on the problem and provide enough information to calibrate our model. At a minimum, simulations should report the final outer radius of the bound material and the final energies of the companion star (as compared to the initial energies).  Better still would be to report asymmetrically averaged density and energy profiles that could then be used as the basis for more self-consistent models of their subsequent evolution.

\acknowledgments

We thank Marc Pinsonneault, Jennifer van Saders, Bill Paxton, Philipp Podsiadlowski, Todd Thompson, Dale Mudd, and Joe
Antognini for discussions and encouragement. We also thank the anonymous referee for their helpful comments and suggestions improving this manuscript.  We acknowledge with thanks the variable star observations from the AAVSO International Database contributed by observers worldwide and used in this research. BJS was supported by a Graduate Research Fellowship from the National Science Foundation.  BJS, CSK and KZS are supported in part by NSF grant AST-0908816. This research has made use of the NASA/IPAC Extragalactic Database (NED) which is operated by the Jet Propulsion Laboratory, California
Institute of Technology, under contract with the National Aeronautics and Space Administration. This research has made use of NASA's Astrophysics Data System Bibliographic Services.

\bibliographystyle{apj}

\end{document}